\documentclass[twocolumn,showpacs,aps,prl,citeautoscript]{revtex4}

\usepackage{graphicx}
\usepackage{dcolumn}
\usepackage{bm}

\begin{document}

\begin{titlepage}

\title{Role of Symmetry in the Transport Properties of Graphene Nanoribbons under Bias}

\author{Zuanyi Li, Haiyun Qian, Jian Wu, Bing-Lin Gu, and Wenhui Duan}\email[Author to whom correspondence should
be addressed. ]{dwh@phys.tsinghua.edu.cn}
\address{Department of Physics, Tsinghua University, Beijing
100084, People's Republic of China}
\date{20 May 2008}

\begin{abstract}

The intrinsic transport properties of zigzag graphene nanoribbons
(ZGNRs) are investigated using first principles calculations. It
is found that although all ZGNRs have similar metallic band
structure, they show distinctly different transport behaviors
under bias voltages, depending on whether they are mirror
symmetric with respect to the midplane between two edges.
Asymmetric ZGNRs behave as conventional conductors with linear
current-voltage dependence, while symmetric ZGNRs exhibit
unexpected very small currents with the presence of a conductance
gap around the Fermi level. This difference
is revealed to arise from different coupling between
the conducting subbands around the Fermi level, which is dependent
on the symmetry of the systems.

\end{abstract}

\maketitle


\end{titlepage}

Along with recent experimental progress in preparing a single layer
of graphite \cite{Novo-S,Berger,Novo-N,Kim-N}, graphene, this
two-dimensional (2D) electronic system has attracted extensive
interest owing to its unusual band structure \cite{Review}. In
particular, graphene can be patterned via standard lithographic
techniques into new quasi-one-dimensional materials
\cite{Berger-S,Kim-gap}, graphene nanoribbons (GNRs), which have
many properties similar to carbon nanotubes, such as energy gap
dependence of widths and crystallographic orientations
\cite{Kim-gap,Gap-Louie-Barone}, and ballistic transport
\cite{White-ballistic}. All these experimental developments and
unique properties of graphene and GNRs provide an exciting
possibility of new nanoelectronics based on graphene and GNRs
\cite{Berger,Yan,White-circuit}.

A feasible approach to realize graphene-based electronics is to
construct the device junctions by connecting GNRs with different
widths and orientations \cite{Yan,White-circuit}. It is thus
necessary to systematically study the transport properties of GNRs
as fundamental and crucial components of graphene-based circuits.
Until now, many efforts have been made in this direction
\cite{White-ballistic,Waka,Peres,Rojas}, and a few intriguing
phenomena were predicted, such as zero-conductance resonance
\cite{Waka-L}, half-metallic conduction \cite{Louie-N}, and valley
filtering \cite{Beenakker-NP}. Most of these theoretical studies
mainly focused on the conductance of GNR junctions at zero bias
voltage ($V_{\rm bias}$) rather than a finite $V_{\rm bias}$.
However, the results of zero bias voltage sometimes can not
represent the overall transport properties of systems, because the
electric field induced by the bias can change the conductance near
the Fermi level \cite{Datta}. Since the electronic devices always
work under a finite bias voltage in practice, it is essential to
gain better understanding of electronic transport properties of
GNRs, especially upon application of a finite bias voltage.

Thus far, in most studies of transport properties of GNRs, all GNRs
with zigzag edges (ZGNRs) are treated as the same type, because they
have very similar metallic electronic properties and transmission
spectra under zero bias voltage
\cite{White-ballistic,Waka,Peres,Rojas}. In this work, however, we
will demonstrate that they can be classified into two groups that
have completely different current-bias-voltage ($I$-$V_{\rm bias}$)
characteristics (but similar band structures) with respect to
whether there exists mirror plane $\sigma$ [Figs. 1(a) and 1(b)].
Asymmetric ZGNRs behave as conventional conductors with one
conductance quantum under bias voltages, whereas symmetric ZGNRs
display very small currents because of the appearance of the
conductance gap depending on the bias voltage. From a two-gates
tight-binding (TB) model, this difference in transport properties is
revealed to originate from different symmetries of ZGNRs, which is
responsible for the different coupling properties between the
subbands. Moreover, it is shown that the current through the
symmetric ZGNRs can be remarkably enhanced by asymmetric edge
terminations. This feature implies that the conduction of ZGNRs can
be controlled by changing their symmetries via external modulations
such as transverse electric fields.

We performed quantum transport calculations of ZGNRs by using an
{\it ab initio} code, \textsc{transiesta-c} \cite{Taylor}, which is
based on real-space, nonequilibrium Green's function formalism and
the density-functional theory, as implemented in the \textsc{siesta}
approach \cite{Soler}. Self-consistent calculations are performed
with a mixing rate of 0.05, and the mesh cutoff of carbon atom is
chosen as 100 Ry.
Our electronic structure calculations were performed using the {\it
ab initio} \textsc{siesta} method \cite{Soler}, which is based on
the DFT within the local density approximation. In the calculations,
structural optimizations were first carried out until atomic forces
converged to 0.02 eV/$\rm{\AA}$. Except where specified, the edges
of ZGNRs are terminated with hydrogen (H) atoms to remove the
dangling bonds. Our calculations confirmed that the energy of the
spin-polarized state (zero-temperature ground state) is $\sim$20 meV
per edge atom lower than that of the spin-unpolarized state
\cite{Louie-N,Lee}. However, the spin-polarized state would become
unstable with respect to the spin-unpolarized state at finite
temperature \cite{Wagner-66} or in the presence of a ballistic
current through the GNRs \cite{White-circuit}. In order to simulate
the experimentally detectable transport behavior under bias
voltages, we focused our study only on the spin-unpolarized metallic
state of ZGNRs. To reveal the physical mechanism behind the results
of {\it ab initio} calculations, we also employed the TB Green
function method \cite{Rojas,Datta,wuapl} to analyze the transmission
spectra with a two-gates model.

\begin{figure}[tbp]
\includegraphics[width=8.2cm]{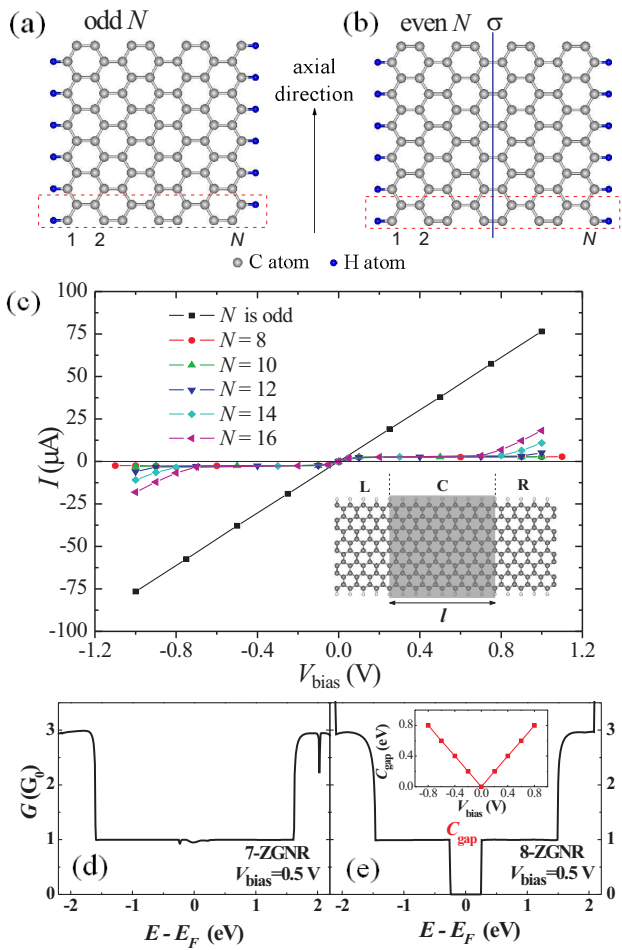}
\caption{(Color online). (a) An asymmetric 7-ZGNR. (b) A symmetric
8-ZGNR, with the mirror plane ($\sigma$) represented by the solid
blue line. The dashed rectangle denotes the unit cell of ZGNRs. (c)
$I$-$V_{\rm bias}$ curves of ZGNR junctions (see the inset) with
different widths. A bias voltage, $V_{\rm bias}$, is applied to the
central region with the length ($l$) of 8 unit cells. (d), (e)
Conductance of 7-ZGNR and 8-ZGNR under $V_{\rm bias}=0.5$ V. The
conductance gap, $C_{\rm gap}$, is shown as a function of $V_{\rm
bias}$ in the inset of (e).} \label{fig01}
\end{figure}

As shown in Figs. 1(a) and 1(b), $N$-ZGNRs \cite{notation} can be
divided into two groups with respect to their symmetries:
\emph{i.e.}, odd(even) $N$ corresponds to asymmetric(symmetric)
ZGNRs. In order to investigate the transport behaviors of ZGNRs
under bias voltages, we carry out extensive first-principle
calculations of $I$-$V_{\rm bias}$ curves by using a two-probe
system [shown in the inset of Fig. 1(c)], where left ($L$) and right
($R$) leads are semi-infinite ZGNRs. Figure 1(c) shows typical
results for ZGNRs with different widths when the length $l$ of the
central region ($C$) is of 8 unit cells. It is interesting to find
that symmetric and asymmetric ZGNRs exhibit completely different
transport characteristics, although they have very similar metallic
band structures. In detail, all asymmetric ZGNRs (namely, $N$ is an
odd number) have linear $I$-$V_{\rm bias}$ curves with the same
slope regardless of the ribbon width and length $l$. As shown in
Fig. 1(d), this result arises from the presence of one conductance
quantum ($G_{0}$) around the Fermi level. On the other hand, all
symmetric ZGNRs (namely, $N$ is an even number) have very small
currents $I$ when $|V_{\rm bias}|$ is smaller than a critical
bias-voltage ($V_{c}$), beyond which the currents begin to increase.
The conductance shown in Fig. 1(e) illustrates that this interesting
phenomenon results from the presence of conductance gap ($C_{\rm
gap}$) around the Fermi level under a certain $V_{\rm bias}$.
Furthermore, this conductance gap increases with increasing $|V_{\rm
bias}|$ but is slightly smaller than $e|V_{\rm bias}|$ [see the
inset in Fig. 1(e)], and thus the current remains small until
$|V_{\rm bias}|$ reaches $V_{c}$.

In general, distinct transport behaviors of two classes of ZGNRs
arise from the characteristics of their electronic structure.
However, as shown in Figs. 2(a) and 2(b), asymmetric and symmetric
ZGNRs exhibit very similar band structures, which agree well with
previous studies \cite{Louie-N,Fujita99}. Around the Fermi level,
there exist two partially flat bands (\emph{i.e.}, edge states)
\cite{Fujita96,Nakada,Fujita99}. As wavevector $k$ deviates from $X$
point, these two edge states mix to form bonding and antibonding
states (\emph{i.e.}, $\pi$ and $\pi^{\ast}$ subbands)
\cite{Waka-L,Fazzio-Boron}. In addition, there exist another two
$V$-type subbands around the Fermi level [see Figs. 2(a) and 2(b)].
Our calculations show that the energy gap between two $V$-type
subbands, $2\Delta$, is inversely proportional to the width of ZGNR,
consistent with previous TB studies \cite{Waka,Beenakker-NP}.

\begin{figure}[tbp]
\includegraphics[width=8cm]{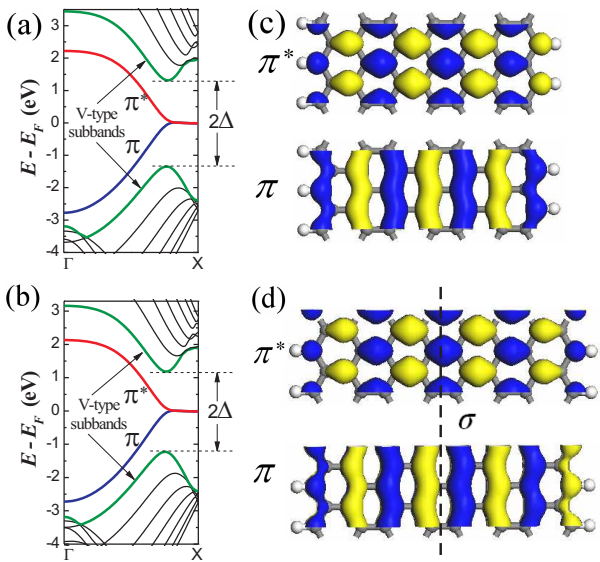}
\caption{(Color online). (a) [(b)] Band structure around the Fermi
level, and (c)[(d)] isosurface plots of the $\Gamma$-point
wavefunctions of $\pi$ and $\pi^{\ast}$ subbands for 7-ZGNR
[8-ZGNR]. In (c) and (d), dark gray (blue) and light gray (yellow)
indicate opposite signs of the wave function.} \label{fig02}
\end{figure}

Clearly, different transport characteristics should come from
something beyond the band structures. We turn to the wave functions
of the $\pi$ and $\pi^{\ast}$ subbands, which are the only bands
near the Fermi level and thus are responsible for the electron
transmission under low bias voltages. As shown in Figs. 2(c) and
2(d), the wave functions of symmetric and asymmetric ZGNRs do
exhibit different characteristics. In accordance with their
symmetric geometry, the $\pi$ ($\pi^{\ast}$) subbands of symmetric
ZGNRs have odd (even) parities under $\sigma$ mirror operation.
While the subbands of asymmetric ZGNRs have no definite parity, due
to the absence of the mirror plane $\sigma$. Then, the problem
becomes why the subband parity affects the electron transmission
under bias voltage.

\begin{figure}[tbp]
\includegraphics[width=8.2cm]{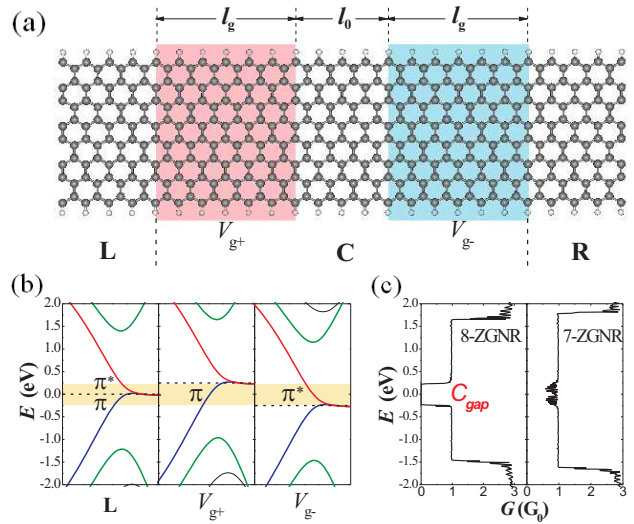}
\caption{(Color online). (a) Schematic configuration of a
two-gates ZGNR model, where positive (negative) gate voltage
$V_{g+}$ ($V_{g-}$) is applied to the red (blue) region. In our
calculations, $l_g$ (the length of gate region) and $l_0$ are
taken to be 50 and 20 unit cells, respectively. (b) Schematic band
structures of the left lead and two gate regions
($V_{g+}$=$-V_{g-}$=0.25 V). The dashed lines represent the Fermi
level of the regions. (c) Transmission spectra of 8-ZGNR and
7-ZGNR in the two-gate model with $V_{g+}$=$-V_{g-}$=0.25 V.}
\label{fig03}
\end{figure}

As is well known, one important effect of the bias voltage $V_{\rm
bias}$ is that the chemical potential will change along the
transport direction. To understand the influence of such change and
subband parity on the transport behaviors, we design an ideal
two-gates model [as shown in Fig. 3(a)] and calculate transmission
spectra using the TB Green function method \cite{Rojas,Datta,wuapl}.
In the system, local gate voltages $V_{g+}$ and $V_{g-}$ are applied
to the two gate regions separated by a buffer region, to simulate
the change of chemical potentials [as shown in Fig. 3(b)]. Figure 3c
shows the calculated transmission spectra of such a two-gate model
for 8-ZGNR and 7-ZGNR with $V_{g+}$=$-V_{g-}$=0.25 V. Evidently,
they display a symmetry-dependent difference, which is the same as
that of ZGNRs under a bias obtained by the {\it ab initio} transport
simulation [Figs. 1(d) and 1(e)]. Then, a clear physical picture can
be drawn from this result. As illustrated in Fig. 3(b), the local
gates shift the Fermi level in the gate regions, and open an energy
window (the yellow region), within which only $\pi$ subband exists
in the $V_{g+}$ region while only $\pi^{\ast}$ subband exists in the
$V_{g-}$ region. For symmetric ZGNRs, the $\pi$ and $\pi^{\ast}$
subbands have opposite $\sigma$ parity, so they can not couple with
each other to contribute to the transmission, and then a conductance
gap $C_{\rm gap}$ appears [the left-hand panel of Fig. 3(c)]. For
asymmetric ZGNRs, however, without the limitation of $\sigma$
parity, a $\pi$ electron in the $V_{g+}$ region may hop to a
$\pi^{\ast}$ state in the $V_{g-}$ region and contributes to the
transmission, leading to the conductance of about 1$G_0$ [the
right-hand panel of Fig. 3(c)].

Then, the $I$-$V_{\rm bias}$ curves of ZGNRs [shown in Fig. 1(c)]
can be understood with the help of the schematic band diagram under
$V_{\rm bias}$ [Fig. 4(a)]. Here, the solid line II represents the
quasi-Fermi level ($F_{q}$) of the system. Lines I and III represent
the critical energies at which $V$-type subbands appear. So only
$\pi$ ($\pi^{\ast}$) subband exists in the energy range between
lines II and III (I). This feature of band structure is important
for the unique transport behaviors of GNRs, and is different from
that of armchair carbon nanotubes where both $\pi$ and $\pi^{\ast}$
subbands (crossing at the Fermi level) span the whole energy range.
The bias voltage makes the chemical potentials of left and right
leads ($\mu_{L}$ and $\mu_{R}$) be separated by $|eV_{\rm bias}|$,
and then only the electrons in the states between $\mu_{L}$ and
$\mu_{R}$ [shaded blue area in Fig. 4(a)] can contribute to the
current \cite{Datta}. When $|eV_{\rm bias}|<\Delta$, all these
incident electrons into the central region are in the $\pi$ state,
so the transmission through the system depends on whether $\pi$
electrons can hop to $\pi^{\ast}$ state or not. For symmetric ZGNRs,
the hopping integral is zero due to $\sigma$ parity limitation, and
consequently the electrons will not flow through the ribbon and the
current is very small. For asymmetric ZGNRs, however, there always
exists conductance near the Fermi level [Fig. 1(d)] due to non-zero
hopping integral between $\pi$ and $\pi^{\ast}$ electrons. On the
other hand, when $|eV_{\rm bias}|>\Delta\approx eV_c$, $V$-type
subbands start to contribute to the conductance, and thus the
currents through symmetric ZGNRs begin to increase.

\begin{figure}[tbp]
\includegraphics[width=8.2cm]{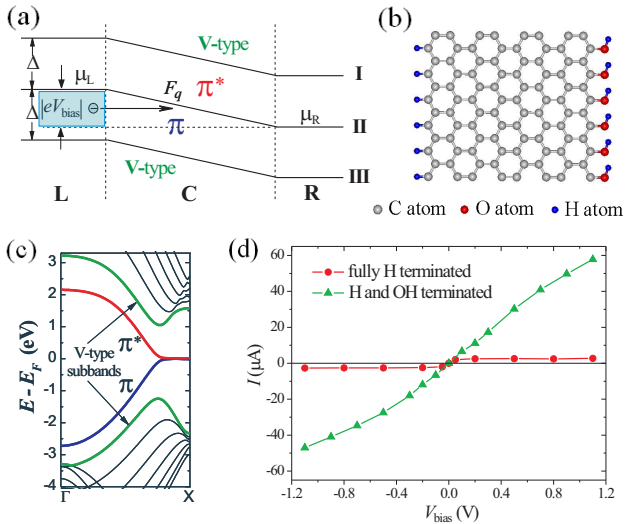}
\caption{(Color online). (a) Schematic band diagram of the
two-probe system under $V_{\rm bias}$. (b) The optimized geometry
and (c) band structure of 8-ZGNR with two edges terminated by H
and OH respectively. (d) $I$-$V_{\rm bias}$ curves of the fully H
terminated 8-ZGNR (red circles), and the H and OH terminated
8-ZGNR shown in (b) (green triangles). $V_{\rm bias}$ is applied
to the central region with the length of 10 unit cells.}
\label{fig04}
\end{figure}

Since the significant deviation from linear $I$-$V_{\rm bias}$
dependence in symmetric ZGNRs (i.e., very small currents under a
finite bias) originates from the symmetry of the system, it can be
expected that breaking the symmetry may enhance the currents. To
confirm this expectation, for original symmetric 8-ZGNR, we use
hydroxyl group (OH) to terminate one edge while using H to terminate
the other edge, and calculate its optimized structure, energy bands,
and $I$-$V_{\rm bias}$ curve as shown in Figs. 4(b)-4(d). Compared
with the results of fully H-terminated 8-ZGNR [Fig. 2(b)], the
energy bands do not change visibly, but the current under bias
voltages increases remarkably. This happens because the asymmetric
edge termination breaks the original mirror symmetry, and
consequently $\pi$ and $\pi^{\ast}$ subbands no longer have definite
$\sigma$ parities. When a bias voltage is applied, $\pi$ and
$\pi^{\ast}$ states can couple with each other and contribute to the
conductance around the Fermi level, which leads to a notably
increasing current. This also implies that the current through ZGNRs
could be tuned by applying a transverse electric field along the
direction perpendicular to the ribbon axes

In conclusion, symmetric and asymmetric ZGNRs are found to have
completely different $I$-$V_{\rm bias}$ characteristics despite
the similarity of their band structure. Asymmetric ZGNRs behave as
conventional conductors with one conductance quantum under bias
voltages, whereas symmetric ZGNRs show very small currents because
of the presence of conductance gap around the Fermi level. It is
the different symmetry of ZGNRs that leads to different coupling
between $\pi$ and $\pi^{\ast}$ subbands and consequently results
in distinct transport behaviors. For symmetric ZGNRs, we further
show that breaking the symmetry of electronic structure can
effectively increase the current through ZGNRs. This might be
useful in the design of electronic devices based on ZGNRs.

This work was supported by the National Natural Science Foundation
of China (Grant Nos. 10325415, 10674077 and 10674078) and the
Ministry of Science and Technology of China (Grant Nos.
2006CB605105 and 2006CB0L0601). The authors thank Qimin Yan and
Zhirong Liu for helpful discussions.


\begin{references}
\bibitem{notation} Herein, $N$ is the number of zigzag chains across the ribbon width,
and hence represents the width of ZGNR [as shown in Figs. 1(a) and
1(b)]. Also see Refs. [25-27] for details of this notation of GNR.
\end{references}
\end{document}